# A NEW METHOD FOR NUCLEAR-REACTION NETWORK ANALYSIS, ELEMENT PRODUCTION IN S-CL REGION, DURING THE HE AND C BURNING IN 25 $M_\odot$ STARS.

*P.Szalanski* [1,4], *I.Padureanu* [2], *M.Stepinski* [1], *Yu.M.Gledenov* [3], *P.V.Sedyshev* [3], *R.Machrafi* [3], *A.Oprea* [3], *D.Aranghel* [2], *J.Marganiec* [1]

[1] *University of Lodz, Poland, e-mail: pjszalan@uni.lodz.pl*
[2] *Horia Hulubei National Institute of Physics and Nuclear Engineering, Bucharest, Romania*
[3] *Joint Institute for Nuclear Research, Dubna, Russia*
[4] *High School of National Economy, Kutno, Poland*

**Abstract**

This paper presents the results of calculation of element overproduction in the S-Cl region during helium and carbon burning in massive stars (25 $M_\odot$). Calculations for a given set of differential equations were performed using integrated mathematical systems. The authors think that this method may be successfully adopted for other physical problems in which solving nuclear-reaction network play an important role, in particular for the problem of transmutation of heavy elements and chemical kinetics.

**Introduction**

Nuclear-reaction networks play an important role in many physical problems. They make it possible to find a simple solution for a natural decay chain of few nuclides, analyse nuclear processes going as activation progresses, obtain a solution for transmutation processes in the reactor core, etc. Many of these problems can be described by simultaneous differential equations of the first degree. An analytical method, the Bateman finite difference, the Runge-Kutta and matrix exponential methods, are generally used to solve these problems [1].

To model a nucleosynthesis process in various star mediums, there are usually used machine optimised dense solvers or matrix specific solvers generated by symbolic processing (for a small network) or generalised sparse solvers, from custom built or software libraries (for a large network) [1]. These methods have rather a good ability to solve a problem presented as a set of differential equations.

In the present work we show that under certain conditions we can successfully use commercial and popular integrated mathematical systems [2,3].

On the other hand, the question concerning the origin of chemical elements in the Universe is of great interest and needs further investigation. Since the work [4] several generations of researchers have contributed new experimental results. However many values required for nucleosynthesis analysis in different stars are known with a great inaccuracy while the others are only theoretically evaluated [5].

Recently, the attention of astrophysicists has focused on the problem of the formation of a rare $^{36}$S isotope (0.02 %) [6]. Initially it was thought that its formation occurred during carbon explosive burning in massive stars (M>10$M_\odot$) [7]. However, a further investigation [8] has not confirmed the approachability of the requirements accepted in [7]. As an alternative astrophysical source of $^{36}$S formation the stage of the hydrostatic burning of helium in these stars was considered [9]. In this case, a $^{36}$S isotope and a series of other elements with a medium or heavy mass may be formed as a result of slow neutron capture (*s*-process). These investigations have continued to present [10-15]. As a neutron source, nuclear reactions in such stars [16,17] and reactions with neutrons have been intensively explored. New experimental results have appeared during the last 10 years [18-22,14,15], and the data obtained in [15,22] differ from theoretical predictions by a factor of 13 and 22. Moreover, in the case of [22] the obtained results have not been completely interpreted from the astrophysical point of view.

**Characteristic of analysed star medium**

Stars with a mass of 25 $M_\odot$ are the most representative sources of nucleosynthesis of massive stars. Generally, a star model includes energy release over the nuclear burning zone, ways of energy transport, related processes inside the star with observable parameters such as radius, mass, surface temperature, and brightness, and a wide network of possible nuclear reactions. Thus, in general it is necessary to investigate the evolution of all parameters in time. However, in many cases a physically motivated network specialisation, where some simplifications are possible, is used to solve local problems. It is clearly seen from the calculation [14] that for He and C burning in massive stars it is possible to average many star parameters in the burning zone instead of performing independent calculations for several tens of different meshes without loss of final results quality.

The burning of helium in stars with a mass of 25 $M_\odot$ occurs in rather stable conditions. The process itself is described in detail in ref. [12]. First of all, in this phase of evolution, during approximately 160 thousand years, in the investigated volume of the star the temperature does not practically vary and it corresponds to $kT$=30 keV. As the considered stars are the stars of the second generation, the initial abundance of isotopes can be accepted to be equal to the solar system abundance [6]. Reactions with charged particles in the stages of hydrostatic burning of H and He do not practically change the abundance of seed elements. To analyse such a medium, in addition to the considered isotopes and the nuclear network including all the dominant reactions (fig. 1) [12,14] an increasing with time neutron density, which does not exceed $10^6$ cm$^{-3}$ at the end of helium burning, must be only taken into account.

During the subsequent stage of convective shell carbon burning the *s*-process nucleosynthesis continues for a relatively short period of 0.66 year although at a higher temperature ($kT$=86 keV) and electron density ($n_e$=3·$10^{28}$ cm$^{-3}$) [23]. These conditions may essentially change the lifetime of unstable nuclides [5,24,25].

**Calculations and results**

To solve such problems, it is necessary to create a linear system of differential equations into which the following may enter:
- an equation (or differential equation) describing the dependence of the neutron density on time;
- differential equations describing the behaviour of the contents of all considered nuclides under activity of possible reactions.

Each term of this equation is responsible for the production or disintegration of the seed nuclide caused by a specified reaction or decay. In our work we use a nuclear reaction-network of 30 nuclei and 47 possible reactions (fig. 1).

The Maxwellian Averaged Cross Section (*MACS*) and decay rates are taken from ref. [26,27,28,15,22] and NACRE database ( http://www-astro.ulb.ac.be ). A complete set of these data can be obtained from Internet at http://astrophysics.fic.uni.lodz.pl/data/isinn-10.pdf . In fig. 2, we present the final calculated abundance for different isotopes during He and C burning; fig. 3 presents the obtained overproduction factors (resulting abundance/seed abundance) for several isotopes [29].

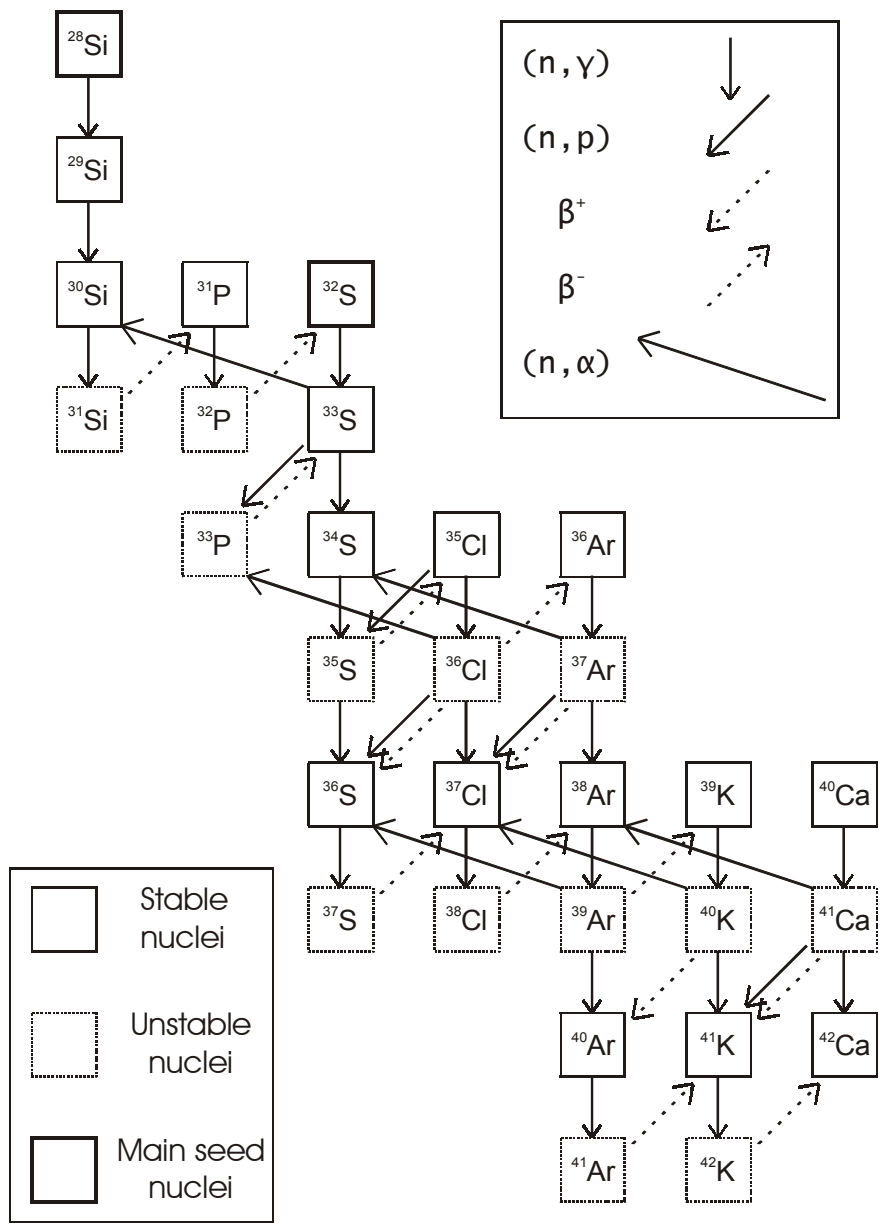

Fig.1. The nuclear-reaction network used in our calculations.

Important branching points of the network were identified as the isotopes $^{33}$S, $^{35}$S, $^{35}$Cl, $^{36}$Cl, and $^{39}$Ar. The corresponding branching ratios were calculated from time integrated reaction flows (for neutron induced reactions see tab. 1) and are shown in tab. 2. The relative importance of each of these reaction sequences to $^{36}$S is shown in tab. 3. During He burning, $^{36}$S production is dominated by the $^{36}$Cl(n,p) reaction, while the $^{35}$S(n,γ) and $^{39}$Ar(n,α) reactions are only important during the C burning stage.

We should also note that in the calculations our aim was not only to obtain the abundance of isotopes but also to explore $^{36}$S formation as a function of the initial abundance of isotopes and possible changes accepted in the calculations of *MACS* quantities. For this purpose two coefficients were introduced [14]:

α - this coefficient shows how the overproduction factor of $^{36}$S isotope varies if one of the *MACS* used increases (+) or decreases (-) two times,

β - this coefficient shows how the overproduction factor of $^{36}$S isotope varies if the initial abundance of one isotope increases (+) or decreases (-) two times.

**Conclusions**

The calculations we have carried out show that the use of integrated mathematical systems allows one to solve physical problems presented as a system of differential equations. In particular, by using our method it is possible to model successfully nucleosynthesis during helium and carbon burning in 25 M$_\odot$ mass stars.

From the point of view of nuclear astrophysics our calculations have shown that the obtained *MACS* in ref. [22] for the $^{37}$Ar(n,p)$^{37}$Cl and $^{37}$Ar(n,α)$^{34}$S reactions do not practically have influence on the formation of $^{36}$S isotopes during He and C burning in 25 M$_\odot$ stars.

Basically, the conclusions of our calculations are in very good agreement with the deductions presented in ref. [14,15] and; in particular, they confirm that the most influential factor in the formation of $^{36}$S isotopes is *MACS* of the $^{34}$S(n,γ)$^{35}$S reaction. Therefore, it would be very useful to study this reaction more precisely as the unique experimental result appeared to be approximately 13 times less than the prediction of the theoretical analysis [27]. Moreover, we have a considerable inaccuracy of the $^{35}$Cl(n,γ)$^{36}$Cl cross section which may have essential influence on the final abundance of the $^{36}$S isotope (see tab. 4);

From tab. 5 it is clear that the yield of the $^{36}$S isotope is very sensitive to the initial abundance of $^{35}$Cl. For this reason it is necessary to carry out neutron nucleosynthesis analysis taking into account available data on $^{35}$Cl content in massive stars and/or to study more accurately possible changes in the initial abundance of the $^{35}$Cl isotope in reactions with charged particles during H and He burning.

In contrast to ref. [14] we conclude that the calculations using updated values of the neutron cross section do not give any considerable overproduction of the formed content of $^{36}$S isotope in the stage of He and C burning in 25 M$_\odot$ stars (see fig. 3). However, at a higher temperature and density other reaction channels with charged particles and γ (an appreciable effect of the reaction with leptons is possible: ν-process) arise and a change in the concentration of nuclides in subsequent stages of burning (if they are implemented, which depends on the considered star model) and in explosive supernovae may be expected. Nevertheless, up to now there has been no reason to look at massive stars as the main source of $^{36}$S isotopes. To complete the analysis of the $^{36}$S isotope formation in nature, it is also necessary to analyse neutron nucleosynthesis in the S-Cl-Ar region using updated neutron cross section values for *AGB* (Asymptotic Giant Branch) stars [15] which are a source of heavy elements (main component of *s*-process). It is necessary to note that star models determining the main parameters of nucleosynthesis are still suffering from a considerable indeterminacy and are permanently explored and investigated. The model from [12] has not yet finally been established, but taking into account the results of [14,15], we have chosen this model for our calculations.

In the present work we do not discuss possible uncertainties due to *MACS* used in the calculations by the reported method. Some information on the problem is available in ref. [30,31].

We hope to continue the modelling of nucleosynthesis using the described method. In particular we plan to extend the method to the star conditions in which during the considered phase of evolution the temperature together with different parameters, such as the density of particles or *MACS,* varies, considerably. Our experience shows that this is quite possible and as a result, it would allow us to apply our method for the modelling of the nucleosynthesis process going in other conditions of interest from the point of view of nuclear astrophysics. Technically, a similar procedure may be applied to other problems, such as neutron activation analysis, transmutation of heavy elements and chemical kinetics.

It is important to note, that the performance of these calculations do not require a long time as one cycle without data storage operations only takes about 20 seconds on a personal computer with an AMD K6-2 (300 MHz) processor and 256 MB of operating memory. Besides, the language of integrated mathematical systems is rather simple and any person who has no wide experience in programming can use it.

The work has been carried out with a partial support of the Centre of Excellence IDRANAP (Bucharest-Magurele) under Contract with the European Commission ICA1-CT-2000-70023.

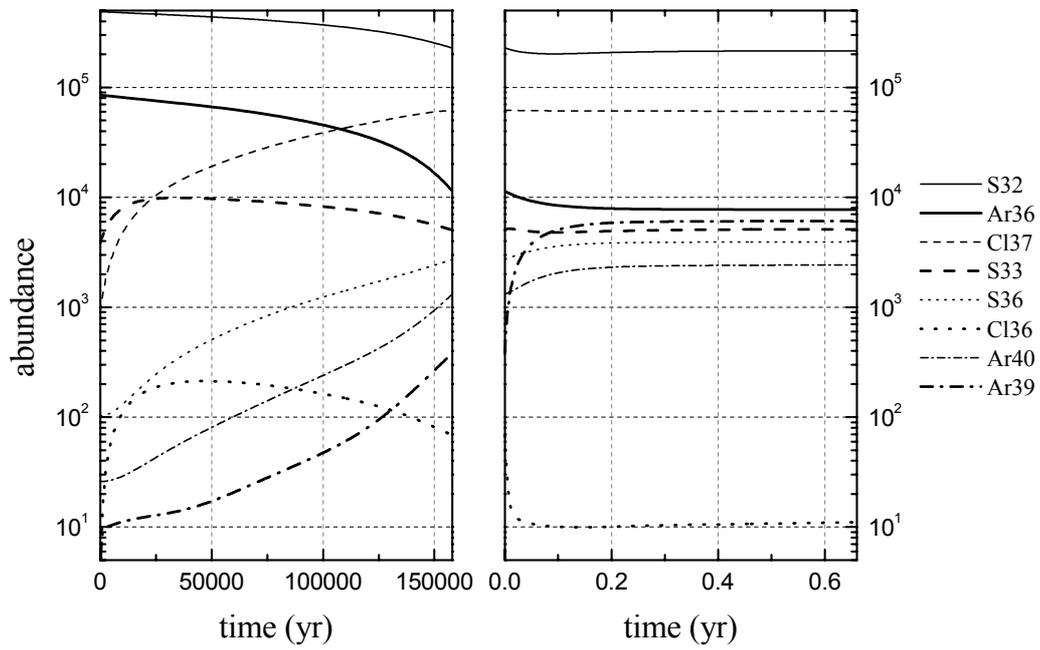

Fig. 2. The time dependence of the isotope abundance during the He and C burning.

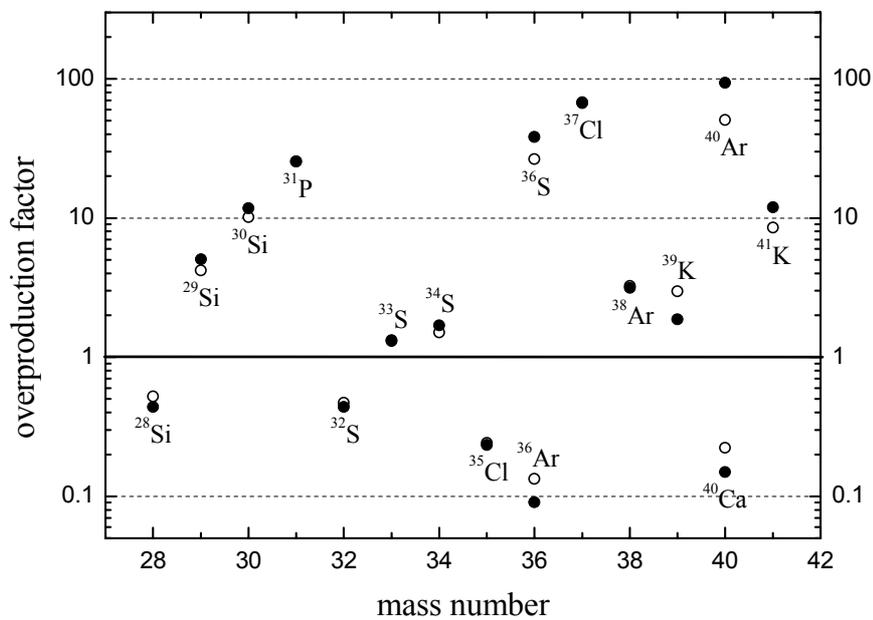

Fig. 3. The resulting overabundances as a function of atomic mass. Open circles indicate the abundances after He burning, while the final abundances after He and C burning are given by filled circles.

Tab. 1. The reaction flow for neutron induced reactions (reactions per $10^6$ Si seed nuclei).

| No. | Reaction type | He burning | C burning | He and C burning |
|---|---|---|---|---|
| 1 | $^{28}$Si(n,γ) | 4.40·10$^5$ | 7.71·10$^4$ | 5.17·10$^5$ |
| 2 | $^{29}$Si(n,γ) | 2.90·10$^5$ | 3.80·10$^4$ | 3.28·10$^5$ |
| 3 | $^{30}$Si(n,γ) | 2.97·10$^5$ | 4.21·10$^4$ | 3.39·10$^5$ |
| 4 | $^{31}$P(n,γ) | 4.32·10$^5$ | 4.14·10$^5$ | 8.47·10$^5$ |
| 5 | $^{32}$S(n,γ) | 3.03·10$^5$ | 5.52·10$^4$ | 3.58·10$^5$ |
| 6 | $^{33}$S(n,γ) | 1.19·10$^4$ | 1.54·10$^3$ | 1.34·10$^4$ |
| 7 | $^{34}$S(n,γ) | 9.04·10$^2$ | 7.70·10$^2$ | 1.67·10$^3$ |
| 8 | $^{35}$S(n,γ) | 5.20·10$^{-3}$ | 5.20·10$^2$ | 5.20·10$^2$ |
| 9 | $^{36}$S(n,γ) | 6.24·10$^1$ | 2.11·10$^1$ | 8.35·10$^1$ |
| 10 | $^{35}$Cl(n,γ) | 3.07·10$^3$ | 2.20·10$^2$ | 3.29·10$^3$ |
| 11 | $^{36}$Cl(n,γ) | 3.41·10$^2$ | 6.63·10$^0$ | 3.47·10$^2$ |
| 12 | $^{37}$Cl(n,γ) | 2.13·10$^4$ | 3.76·10$^3$ | 2.51·10$^3$ |
| 13 | $^{36}$Ar(n,γ) | 7.37·10$^4$ | 3.67·10$^3$ | 7.73·10$^4$ |
| 14 | $^{37}$Ar(n,γ) | 1.30·10$^{-1}$ | 2.27·10$^2$ | 2.27·10$^2$ |
| 15 | $^{38}$Ar(n,γ) | 2.55·10$^4$ | 7.32·10$^3$ | 3.28·10$^4$ |
| 16 | $^{39}$Ar(n,γ) | 2.48·10$^2$ | 1.05·10$^3$ | 1.30·10$^3$ |
| 17 | $^{40}$Ar(n,γ) | 3.05·10$^2$ | 3.48·10$^2$ | 6.53·10$^2$ |
| 18 | $^{39}$K(n,γ) | 1.79·10$^4$ | 3.97·10$^3$ | 2.18·10$^4$ |
| 19 | $^{40}$K(n,γ) | 6.63·10$^3$ | 1.42·10$^3$ | 8.05·10$^3$ |
| 20 | $^{41}$K(n,γ) | 6.84·10$^3$ | 2.15·10$^3$ | 8.98·10$^3$ |
| 21 | $^{40}$Ca(n,γ) | 4.60·10$^4$ | 4.40·10$^3$ | 5.04·10$^4$ |
| 22 | $^{41}$Ca(n,γ) | 3.90·10$^3$ | 9.20·10$^2$ | 4.82·10$^3$ |
| 23 | $^{33}$S(n,p) | 7.70·10$^2$ | 8.50·10$^2$ | 1.62·10$^4$ |
| 24 | $^{35}$Cl(n,p) | 5.75·10$^2$ | 6.11·10$^1$ | 6.36·10$^2$ |
| 25 | $^{36}$Cl(n,p) | 2.59·10$^3$ | 2.67·10$^2$ | 2.85·10$^3$ |
| 26 | $^{37}$Ar(n,p) | 7.04·10$^{-1}$ | 9.94·10$^1$ | 1.00·10$^2$ |
| 27 | $^{40}$K(n,p) | 1.35·10$^3$ | 4.11·10$^2$ | 1.76·10$^3$ |
| 28 | $^{41}$Ca(n,p) | 1.08·10$^3$ | 1.23·10$^3$ | 2.31·10$^3$ |
| 29 | $^{33}$S(n,α) | 2.90·10$^5$ | 5.36·10$^4$ | 3.44·10$^5$ |
| 30 | $^{36}$Cl(n,α) | 2.56·10$^1$ | 2.93·10$^0$ | 2.85·10$^1$ |
| 31 | $^{37}$Ar(n,α) | 1.91·10$^1$ | 3.29·10$^3$ | 3.31·10$^3$ |
| 32 | $^{39}$Ar(n,α) | 9.91·10$^1$ | 4.44·10$^2$ | 5.43·10$^2$ |
| 33 | $^{40}$K(n,α) | 8.34·10$^3$ | 2.55·10$^3$ | 1.09·10$^4$ |
| 34 | $^{41}$Ca(n,α) | 4.00·10$^4$ | 1.74·10$^4$ | 4.17·10$^4$ |

Tab. 5. The sensitivity of the overproduction factor of $^{36}$S isotope as a function of the initial abundance of seed nuclei (results after He and C burning).

| No. | Nuclide | β(+) | β(-) |
|---|---|---|---|
| 1 | $^{35}$Cl | 1.774 | 0.611 |
| 2 | $^{34}$S | 1.121 | 0.938 |
| 3 | $^{32}$S | 1.027 | 0.985 |
| 4 | $^{40}$Ca | 1.018 | 0.989 |
| 5 | $^{38}$Ar | 1.008 | 0.995 |

Tab. 2. The branching points in the reaction flow.

| Isotope | Reaction channel | Reaction flow (%) | |
|---|---|---|---|
| | | He burning | C burning |
| $^{33}$S | (n,α) | 95.83 | 95.73 |
| | (n,γ) | 3.92 | 2.75 |
| | (n,p) | 0.25 | 1.52 |
| $^{35}$S | (β⁻) | 100.00 | 66.86 |
| | (n,γ) | 3.5·10⁻⁴ | 33.14 |
| $^{35}$Cl | (n,γ) | 84.2 | 78.24 |
| | (n,p) | 15.8 | 21.76 |
| $^{36}$Cl | (n,p) | 86.09 | 96.55 |
| | (n,γ) | 11.35 | 2.39 |
| | (β⁻) | 1.70 | 1.8·10⁻¹¹ |
| | (n,α) | 0.85 | 1.06 |
| $^{39}$Ar | (β⁻) | 98.62 | 7.82 |
| | (n,γ) | 0.99 | 64.74 |
| | (n,α) | 0.39 | 27.44 |

Tab. 3. The *s*-process contributions to $^{36}$S production.

| Reaction | Contributions in % | | |
|---|---|---|---|
| | He burning | C burning | He+C burning |
| $^{36}$Cl(n,p) | 94.51 | 21.70 | 72.82 |
| $^{39}$Ar(n,α) | 3.62 | 36.06 | 13.87 |
| $^{35}$S(n,γ) | 1.9·10⁻⁴ | 42.23 | 13.28 |
| $^{36}$Cl(β⁻) | 1.87 | 1.9·10⁻⁵ | 0.03 |
| for total production | 68.95 | 31.05 | 100.00 |

Tab. 4. The sensitivity of the overproduction factor of $^{36}$S isotope versus *MACS* (results after He and C burning).

| No. | Reaction type | α(+) | α(−) |
|---|---|---|---|
| 1 | $^{35}$Cl(n,γ)$^{36}$Cl | 1.132 | 0.752 |
| 2 | $^{34}$S(n,γ)$^{35}$S | 1.147 | 0.924 |
| 3 | $^{36}$Cl(n,p)$^{36}$S | 1.078 | 0.873 |
| 4 | $^{36}$Cl(n,γ)$^{37}$Cl | 0.906 | 1.054 |
| 5 | $^{39}$Ar(n,α)$^{36}$S | 1.035 | 0.981 |
| 6 | $^{33}$S(n,γ)$^{34}$S | 1.026 | 0.985 |
| 7 | $^{33}$S(n,α)$^{30}$Si | 0.985 | 1.023 |
| 8 | $^{36}$S(n,γ)$^{37}$S | 0.977 | 1.010 |
| 9 | $^{32}$S(n,γ)$^{33}$S | 1.017 | 0.987 |
| 10 | $^{40}$Ca(n,γ)$^{41}$Ca | 1.005 | 0.992 |